# Versioned external-memory dictionaries with optimal query/update tradeoffs

Andrew Byde
Acunu Ltd
andrew@acunu.com

Andy Twigg
Acunu Ltd
andy@acunu.com

## ABSTRACT
External-memory dictionaries are a fundamental data structure in file systems and databases. Versioned (or fully-persistent) dictionaries have an associated version tree where queries can be performed at any version, updates can be performed on leaf versions, and any version can be 'cloned' by adding a child. Various query/update tradeoffs are known for unversioned dictionaries, many of them with matching upper and lower bounds. No fully-versioned external-memory dictionaries are known with optimal space/query/update tradeoffs. In particular, no versioned constructions are known that offer updates in $o(1)$ I/Os using $O(N)$ space. We present the first cache-oblivious and cache-aware constructions that achieve a wide range of optimal points on this tradeoff.

## General Terms
Cache-oblivious algorithms, External-memory algorithms, Versioned data structures

## 1. INTRODUCTION
We study tradeoffs between space, query cost and update cost for versioned external-memory dictionaries. A versioned dictionary stores keys and their values with an associated version tree, and supports the following operations:

- `update(key, value, version)`: associate value to the key in the specified leaf version

- `query(start, end, version)`: return every key in the range [start,end] together with the value written in the closest ancestor to version

- `clone(version)`: create a new version as a child of the specified version

A versioned dictionary can be thought of as efficiently implementing the union of many dictionaries: the 'live' keys at version $v$ are the union of all the keys in ancestor versions, where if a key appears more than once, its closest ancestor takes precedence. If the structure supports arbitrary version trees, then we call it (fully-)versioned; if it supports only linear version trees, we call it partially-versioned. We are interested in fully-versioned structures. We use $N$ to denote the total number of keys written; for a version $v$, we use $N_v$ to denote the number of keys that are live at $v$.

The B-tree [2] is the classic external memory dictionary. More recently, data structures that achieve a wide range of query/update tradeoffs have been discovered, in particular, those that offer updates in $o(1)$ I/Os while increasing query cost slightly are of great practical interest.

We aim to answer the following open questions:

1. Can one achieve optimal $O(N)$ space with the same query/update bounds as a CoW B-tree?

2. Can one achieve other points on the tradeoff curve?

3. Can these be achieved in both the DAM and CO models?

Even ignoring updates, it is already difficult to efficiently answer range queries with little space. For deep version trees, many keys in the range may not have been updated since the root version, while some may have been updated many times since then. It is easy to see that some elements must be replicated many times for range queries to be asymptotically optimal – a construction that achieves this while balancing asymptotically optimal space, query and update costs is our main contribution.

As a warm-up, consider the following two naive implementations: keeping a B-tree of the latest key for each version gives excellent query performance, but at the expense of space and update costs. In contrast, keeping a single B-tree with elements ordered by (key,version) uses optimal $O(N)$ space but a small range query may be forced to scan all the elements in $O(N/B)$ I/Os.

### 1.1 Unversioned query/update tradeoffs
The B-tree [2] is the classic external-memory dictionary. The B-tree is typically analyzed in the disk access machine (DAM) model [14]; this assumes an internal memory of size

$M$ and an arbitrarily large external memory where each IO can read or write a block of $B$ elements. An $N$-node B-tree supports updates in $O(\log_B N)$ such I/Os and range queries returning $Z$ elements in $O(\log_B N + Z/B)$ I/Os. An important characteristic of the B-tree is that it is optimal for searching within the DAM model.

It has been observed that there is a tradeoff between query and update performance, and that B-trees achieve only one point on this tradeoff. The buffered-repository tree (BRT) [9] supports updates in amortized $O(\log N/B)$ I/Os and queries in $O(\log N)$ I/Os. Hence, searches are slower in the BRT than in the B-tree, whereas updates are significantly faster. More generally, the $B^\varepsilon$-tree of Brodal and Fagerberg [8] supports a large part of this tradeoff: for $0 \leq \varepsilon \leq 1$, the $B^\varepsilon$-tree supports updates in amortized $O(\frac{\log_{B^\varepsilon+1} N}{B^{1-\varepsilon}})$ I/Os and searches in $O(\log_{B^\varepsilon+1} N)$ I/Os. Thus, when $\varepsilon = 1$ it matches the performance of a B-tree, and when $\varepsilon = 0$, it matches the performance of a BRT. An interesting intermediate point is when $\varepsilon = 1/2$; then searches are slower by a factor of roughly 2, but updates are roughly $\sqrt{B}/2$ faster than a B-tree.

Similar results are known for the cache-oblivious (CO) model [12]. The CO model is similar to the DAM model, except that the block size $B$ is unknown to the algorithm and cannot be used as a tuning parameter. The COLA of Bender et al. [5] achieves the same tradeoffs as the BRT in the CO model. More recently, Brodal et al. [7] presented a CO algorithm that achieves the same range of tradeoffs as the $B^\varepsilon$-tree. It is worth noting that all these schemes achieve the optimal $O(N)$ space bound – it has not been necessary to use more space in order to achieve the tradeoffs in either model.

## 1.2 Versioned query/update tradeoffs

No similar tradeoffs are known for versioned dictionaries, either in the DAM or CO models. In fact, matching these bounds in the CO model with fast updates is impossible – Afshani et al. [1] showed that any partially-versioned CO dictionary supporting range queries in $O((\log_B N)^c(1+Z/B))$ I/Os for any $c > 0$ must use $\Omega(N(\log \log N)^\varepsilon)$ space, where $\varepsilon > 0$ depends on $c$. In their model, every update creates a new version (hence there are $N$ versions). By contrast, in our model, we explicitly track a version tree $V$ and new versions are created with a clone operation (and we assume that the version tree can fit entirely in memory; thus, it is perhaps more appropriate to describe our solution as 'semi-external memory').

The classic versioned analogue of the B-tree is the copy-on-write (CoW) B-tree [10], which is based on the 'path-copying' technique originally presented by Driscoll et al. [11] for making internal-memory data structures fully-persistent, but it does not apply efficiently to external-memory structures. It supports updates to version $v$ in $O(\log_B N_v)$ I/Os and range queries of size $Z$ in $O(\log_B N_v + Z/B)$ I/Os. Clearly, these query bounds are the best we can hope for, since $O(\log_B N_v)$ is the bound we would get if we were to isolate all the keys accessible from version $v$ and store them in a B-tree. This data structure is fundamental to every NetApp filer [13], the ZFS file system [6], and in numerous file systems and databases. The basic idea is to use a B-tree with many roots, one for each version. A lookup proceeds as in a B-tree, starting from the appropriate root. An update to key $k$ at version $v$ goes as follows; if there is a root node for $v$, perform a regular B-tree update for $k$ starting at that root node; otherwise, find the root node for $v$'s parent version and perform a regular B-tree lookup for $k$ to find the correct leaf node, then duplicate this entire root-to-leaf path, and finally set the root node of this path as the root node for version $v$.

The CoW B-tree has two major problems that we seek to address: first, it is not space-optimal – in general, each update may cause a new path to be written, giving $\Theta(NB\log_B N)$ space[1] and second, it does not offer any update/query tradeoffs. The 'multiversion B-tree' (MVBT) of Becker et al. [3] achieves $O(\log_B N_v)$ I/Os for updates and queries with $O(N)$ space, but is only partially-versioned and does not support any other tradeoffs.

## 1.3 Our results

We present the first fully-versioned dictionaries that achieve optimal $O(N)$ space, and optimal query/update tradeoffs in both the DAM and CO models. One can see them as fully-versioned analogues of the $B^\varepsilon$-tree [8] and the COLA [5] in the DAM and CO models respectively.

In the DAM model, we present an external-memory versioned dictionary using space $O(N)$ that supports updates to version $v$ in amortized $O(\frac{\log_B N_v}{\varepsilon B^{1-\varepsilon}})$ I/Os and supports range queries of size $Z$ in worst-case $O(\frac{\log_B N_v}{\varepsilon} + \frac{Z}{B})$ I/Os.

In the CO model, we present a cache-oblivious external-memory versioned dictionary that uses space $O(N)$ and supports updates to version $v$ in amortized $O(\log N_v/B)$ I/Os, and range queries at version $v$ returning $Z$ elements in amortized $O(\log N + Z/B)$ I/Os. We can deamortize the structure so that updates run in worst-case $O(\log N_v)$ I/Os (with the same amortized bound), and point queries at version $v$ run in worst-case $O(\log N_v)$ I/Os. Similarly to Bender et al. [5], with knowledge of $B$ ('cache-aware'), the data structure can support updates to version $v$ in amortized $O(\frac{\log_B N_v}{\varepsilon B^{1-\varepsilon}})$ I/Os, and range queries of size $Z$ in amortized $O(\frac{\log_B N_v}{\varepsilon} + \frac{Z}{B})$ I/Os.

Our results leave open two problems: first, fully deamortizing range queries in the CO dictionary, and second, achieving the $\varepsilon$-dependent bounds without knowledge of $B$.

## 2. PRELIMINARIES
## 2.1 Key and Version Ordering

We often discuss ordering elements lexicographically by key and version ('*kv order*'). Keys are assumed to have a natural total ordering, so we shall describe the version ordering. The versions are nodes in a version tree, so we have the ancestor partial order $\preceq$ – we write $x \preceq y$ to mean '$x$ is an ancestor of $y$'. We say that versions $x, y$ are *comparable* if either $x \preceq y$ or $x \succeq y$. For *kv* order, we allow any total order consistent with $\succeq$ in the sense that every version $v$ occurs after all descendants $w \succeq v$. Order-ing versions descending by their DFS number satisfies this, with the advantage that ancestorship can be tested in $O(1)$ time: let the interval $I(v) = [\text{DFS}(v), \max_{w \succeq v} \text{DFS}(w)]$, then $v \preceq w \iff \text{DFS}(w) \in I(v)$. As the version tree changes, we can use an efficient renumbering scheme to retain integer DFS values, such as in the order maintenance problem [4].

---
[1]Typically, $B$ is thousands in practice and $\log_B N < 5$.

## 2.2 Definitions
Consider a set of elements $A$ and versions $V$. An element $(k, v)$ is a *lead* element (at $v$) if $v \in V$. Define $lead(A, v)$ as the total number of lead elements at $v$ in $A$ and $lead(A, V) = \sum_{v \in V} lead(A, v)$. The *lead-below* count is the total lead at versions descendent from $v$, i.e. $lead\_below(v) = \sum_{x \succeq v} lead(v)$.

An element $(k, x)$ is said to be *live* (or *accessible*) at version $v$ in $A$ if $x \preceq v$ and $k$ has not been rewritten between $x$ and $v$, i.e. there is no other element $(k, y) \in A$ with $x \prec y \preceq v$. Let $live(A, v)$ be the total number of elements of $A$ that are live at $v$. Note that if $v \preceq w$ then $live(v) \leq live(w)$. Also

$$live(v) \leq live(parent(v)) + lead(v), \qquad (1)$$

with the difference between right and left-hand sides being equal to the number of keys $k$ which appear in both versions $v$ and $parent(v)$. We use $N$ to denote the total number of keys written; for a version $v$, we use $N_v$ to denote the number of keys that are live at $v$, i.e. the number of distinct keys written in ancestor versions of $v$ (each key is live at least once, so $\sum_v N_v \geq N$).

We assume that keys and values (which could be pointers to data or real data) are all of fixed size.

## 3. A CACHE-OBLIVIOUS VERSIONED B-TREE
In this section we present a cache-oblivious versioned B-tree, which we refer to as a *stratified doubling array* (SDA). It contains a collection of arrays of key-version-value tuples, arranged into levels, with 'forward pointers' to facilitate searching. Arrays in level $l$ are roughly twice as large as arrays in level $l-1$, hence 'doubling', and have disjoint sets of versions associated to them, hence 'stratified' in version space.

The basic idea is to store arrays of $kv$-ordered elements, as in the COLA of Bender et al. [5], except that we apply a version split process, similar to the one employed in the versioned B-tree, albeit more complex, in order to avoid arrays containing too few elements from some version (we call this a 'density' property). The result is that each level may have several arrays, tagged with disjoint sets of versions that indicate which should be used.

### 3.1 Arrays
An *array* $(A, V)$ contains a set $A$ of *entries* $(k, v, x)$ where $k$ is a key, $v$ is a version, and $x$ is either a data value or a forward pointer containing an array index (the array into which it indexes will become clear from the context later), ordered by $(k, v)$. The set $V$ is a set of 'valid versions' that will be used for lookups and merges between various arrays. Each array also contains a pointer to a unique 'next array', identifying the array, if any, into which its forward pointers point. Arrays implement the following operations:

- `search(k,v,[lb],[ub])`: search for a $(k, v)$ pair, within optional lower and upper bounds. It returns the index of a least upper bound $y$ for $(k, v)$ in the k-v order, and the destinations of the two closest forward pointers either side of $y$.

- `iterate(loc)`: provides an iterator over elements starting from index `loc`.

- `append(k,v,x)`: appends the entry to the end of the array, returning its location.

### 3.2 Definitions
For a version $v$, the *density* of version $v$ in $A$ is $\delta(A, v) = live(A, v)/|A|$. We say that a version $v$ is *dense* in $A$ if $\delta(A, v) \geq 1/3$, and that an array $(A, V)$ is dense if every $v \in V$ is dense in $A$. Note that if $v$ is dense in $(A, V)$ then every descendant version is also dense there.

Given a non-empty set of versions $V$, we say a version $v$ is an *orphan* of $V$ if it has no strict ancestor in $V$. We say the array $(A, V)$ is a *stratum* if the orphans of $V$ are all siblings – they have the same parent, not in $V$, which we write without ambiguity as $parent(V)$;

For a version $v$ and set of versions $V$, let $T_V[v] = \{w \in V : v \preceq w\}$ be the subtree of $V$ rooted at $v$. For $W \subset V$ a set of versions and $A$ an array, define the *split* of $A$ with respect to $W$ to be the set of all entries live in any version in $W$: $\lambda(A, W) = \{(k, x) \in A : (k, x) \text{ is live at some } v \in W\}$, i.e. the set of all keys live in any version in $W$. For $W$ a stratum with orphans $w_i$ having common parent $p$, define

$$\begin{aligned} \texttt{arr\_size}(A, W) &:= live(A, p) + lead(A, W) \\ &= live(A, p) + \sum_i lead\_below(A, w_i) \end{aligned} \qquad (2)$$

As in (1), $|\lambda_T(A, W)| \leq \texttt{arr\_size}(A, W)$ with the difference being those keys live in the parent version but over-written in all orphans of $W$.

As a special case, when $W = T_V[v]$ for some version $v \in V$, define $\lambda_T(A, V, v) = \lambda(A, T_V[v])$, and as usual where $A$ and $V$ are clear, we write $T[v]$ and $\lambda_T(v)$ for the set of versions and corresponding split respectively. Note that $lead(A, T_V[v]) = lead\_below(A, v)$.

A *version split* of an array $(A, V)$ gives a set of strata $\{(A_i, V_i)\}_i$ such that $A = \cup_i A_i$, and $V = \cup_i V_i$, and $V_i$ are mutually disjoint.

### 3.3 Levels
As previously mentioned, an SDA keeps $(k, v, x)$ tuples in arrays arranged into levels. Each level $l \geq 0$ contains a set of arrays $(A_i^l, V_i^l)$ with disjoint sets of valid versions. We keep in memory a map from version to the array in which it is valid – if such a thing exists. We also keep track of the subset of those versions (which we call 'real') for which is at least one lead key in the array where $v$ is valid.

### 3.3.1 Promotion Conditions

Before describing invariant properties of levels, we introduce the following logical conditions on arrays and versions:

- We try to ensure that arrays at level $l$ have sizes in the range $2^l \leq |A| < 2^{l+1}$; we refer to these size conditions as (P-min-size)$_{l,A}$ : $|A| \geq 2^l$, and its contrary (P-max-size)$_{l,A} = \neg$(P-min-size)$_{l+1,A}$.

- It will be important for all arrays, both those in a level and those being promoted, to have a suitably large number of keys live in each version; such a lower bound on the number of live keys clearly implies a density constraint given the size constraints above. Formally we'll refer to (P-live)$_{l,v}$ : $live(v) \geq 2^l/3$.

- Likewise it often turns out to be important that no strict ancestor of an array has so many keys live. The intuition here is that we want to ensure that there are not too many copied keys in an array so that merges can be 'paid' for by insertion of lead keys. The constraint we'll refer to is simply the contradiction of P-live: (P-plive)$_{l,V} = \neg$(P-live)$_{l+1,parent(V)}$.

- In order to be able to argue that the amount of merge work done is bounded by a linear function of the number of keys inserted, we will insist on a lower bound on the number of lead keys in each promoted array: (P-lead)$_{l,v}$ : $lead\_below(v) \geq 2^{l+1}/3$;

- Putting it together, when searching for arrays to promote we will look for a version $v$ whose subtree satisfies the conjunction of the above properties: (P-prom)$_{l,v}$ = (P-live)$_{l+1,v}$ ∧ (P-lead)$_{l+1,v}$ ∧ (P-min-size)$_{l+1,\lambda_T(v)}$.

The following properties hold for every array $(A, V)$ at level $l$ with at least one real version (i.e. version with lead > 0):

- (L-dense) $A$ is dense for all versions in $V$.

- (L-size) $A$ is not too big: (P-max-size)$_{l,A}$;

- (L-live) $A$ has a minimal number of keys live for every version: (P-live)$_{l,v}$ for all $v \in V$;

- (L-plive) The parent of $A$ has few keys live: (P-plive)$_{l,V}$;

- (L-no-prom) There are no versions in $A$ which are 'promotable' in the sense used above: $\forall v \in V, \neg$ (P-prom)$_{l+1,v}$

- (L-edge) If there are enough keys in version $v$ to justify promotion (ignoring the lead requirement) then no descendant can have reached a strictly higher level yet: $\forall v \in V,$ (P-live)$_{l+1,v} \implies$ no level $l' > l$ contains a key in a version strictly descendant from $v$.

We will show how to maintain these properties under updates.

## 3.4 Updates and promotions

Promotion is the process by which an array is moved up from one level to the next and merged with an existing array there. The update(k,v,x) operation is itself considered to be a promotion – the promotion of the singleton array $A = [(k,v,x)]$ with valid version set $V = \{v\}$, to level 0. In general, the only sort of array $(A, V)$ that will be promoted from one level to the next is of the form $\lambda_T(v)$ for a suitable $v$. The properties we are interested in for $(A, V)$ are:

- $V$ has a unique orphan $v$, which makes the choice of array with which to merge simple;

- $A$ has a large fraction of lead keys, which allows us to account for the cost of merging and splitting;

- versions obey a density property, which in conjunction with density for existing arrays in the target level allows us to maintain (L-live).

We will show that the result of this merge can be 'version split' into new arrays which are either suitable to remain at level $l$, satisfying the level requirements above, or are suitable to be promoted to the next level $l+1$. Often the result of the merge need not be split at all, or can be promoted in its entirety. To be precise, an array $(A, V)$ promoted to level $l$ will satisfy the following 'promotion conditions':

- (P-orphan): $V$ has a unique orphan $v$;

- (P-non-trivial)$_v$ : $lead(v) > 0$;

- (P-max-size)$_{l,A}$ ($A$ is not too big);

- (P-plive)$_{l,V}$ ($A$'s parent has few live keys);

- (P-prom)$_{l,v}$ (... but $A$ itself is promotable: it has large enough live and lead counts, and is big enough);

- (P-edge): no level $l' \geq l$ contains a key in a version strictly descendant from $v$.

Note that a single insert into level 0 satisfies these conditions.

## 3.5 Algorithm Overview

The choice of which array at level $l$ to merge $(A, V)$ with orphan $v$ into is simple: if $v$ is registered to some array $A'$, then merge into that; else, if the next array to which forward pointers in $A$ point is at level $l$, then merge into that; otherwise, there is no suitable array: $A' = \emptyset$.

Our general approach is to first calculate an appropriate set of output version sets, based on lead and live statistics for the input arrays $A$ and $A'$; to each such version set we will associate an output array, initially empty; then we will iterate over the contents of the input arrays in $k, v$-order, appending each entry to appropriate output arrays.

This process is I/O efficient, since it requires one complete sequential read across each input array, and sequential writes to each output array. Importantly, for practical implementations, with sufficient prefetching and buffering, it can take advantage of sequential I/O. After the output arrays are generated, forward pointer sample arrays will be back-propagated down towards level 0, and at most one will be promoted up to the level above. Thus the merge operation can be decomposed into the following phases:

1. seek a **promotable** version set, and remove it if one is found;
2. perform a **version split** of the remainder;
3. **execute** the resulting promotion and version split; and
4. **back-propagate** forward pointer arrays.

We now describe each of these phases in detail.

### 3.5.1 Finding a Promotable Version
We say a version $w \in V''$ is **promotable** if $\lambda_T(w)$ obeys the promotion conditions for level $l+1$, most importantly (P-prom)$_{l+1,\lambda_T(w)}$. Using the statistics of the merged array $(A'', V'')$, we search for the promotable version $w$ for which $|\lambda_T(w)|$ is the largest possible. This can be done by searching recursively down through $V''$, starting with whichever orphan $z$ of $V''$ is ancestral to $w$. Note that once the (P-lead) or (P-min-size) conditions fail, the whole search fails, since both are non-increasing down the tree.

---
**Algorithm 1** find_promotable($W, w$)
---
**Require:** A threshold size $M$
1: **if** $|\lambda_T(w)| < M$ **or** $lead\_below(w) < 2M/3$ **then**
2:    **return** $null$
3: **else if** $lead(w) > 0$ **and** $live(w) \geq M/3$ **then**
4:    **return** $w$
5: **else**
6:    **for** $u$ in the children of $w$ **do**
7:      let $u' = $ find_promotable($W, u$)
8:      **if** $u'$ is not $null$ **then**
9:         **return** $u'$
10:      **end if**
11:    **end for**
12:    **return** $null$
13: **end if**

---

Pseudo-code is given in Algorithm 1. We search for a promotable version $w = $ find_promotable($V'', z$) with $M = 2^{l+1}$. If $w$ is $null$, then we proceed to the version split phase using $(A'', V'')$, otherwise we remove the subtree rooted at $w$ from $V''$ before proceeding, using the suitably diminished counts for $lead\_below(\cdot)$ on the remaining versions. Both the elements extracted by find_promotable and the remainder satisfy the desired properties.

LEMMA 1. *Suppose $(A'', V'')$ is the result of merging a promoted array $(A, V)$ into an existing array $A'$ at level $l$. Let $z$ be the unique orphan of $V''$ ancestral to the orphan of $V$, and $w = $ find_promotable$(V'', z) \neq null$. Then (1) $(A_P, V_P) = \lambda_T(A'', V'', w)$ obeys the promotion conditions at level $l+1$; and (2) the remainder $(A_R, V_R)$ of $(A'', V'')$ after $A_P$ is removed, obeys (L-no-prom).*

PROOF. First note that the algorithm guarantees (P-non-trivial)$_w$ and (P-prom)$_{l+1,w}$, and that (P-orphan) is obvious.

The version $w$ cannot be a strict descendant of $v$, since in this case (P-edge) for $A$ would imply that $A_P \subset A$; (P-min-size) cannot hold at level $l+1$ for a subarray of $A$, for which (P-size) holds at level $l$. Likewise $w$ cannot be unrelated to $v$, since in that case $\lambda_T(A'', w) = \lambda_T(A', w)$ and so (P-prom)$_{l+1,w}$ is in contradiction to (L-no-promote). So $w$ is a weak ancestor of $v$.

If $w = v$ then (P-edge) at level $l$ for $A$ implies (P-edge) at level $l+1$, since it's a weaker constraint; if $w \prec v$ then the number of keys live for $w$ has not increased as a result of promotion, and so (P-live)$_{l+1,w}$ must have held prior to the promotion of $A$; the necessary (P-edge) constraint is then a consequence of (L-edge) prior to promotion.

The only remaining condition which remains to be checked is (P-max-size)$_{l+1,A_P}$ – that $A_P$ it is not too large. However, $A_P \subset A''$ and $|A''| \leq |A| + |A'| < 2^{l+2}$, by (L-size) and (P-max-size) for $A'$ and $A$ respectively.

The second part of the lemma follows from the fact that find_promotable finds the *oldest* promotable version: suppose $w' \in V_R$ has at least $2^{l+1}/3$ keys live in $A_R$. We have $w' \not\prec w$ since otherwise find_promotable would have chosen $w'$ before reaching $w$ in the search ((P-lead)$_{l+1,w} \implies$ (P-lead)$_{l+1,w'}$ and likewise for (P-min-size)). We have $w' \not\succeq w$ because no such versions are in $V_R$. Therefore, $w'$ is incomparable to $w$ and also $v$, since $w \preceq v$. But the version statistics of such versions are unaffected in the merge, so (L-no-promote) must hold post-merge since it held for $A'$ before. □

### 3.5.2 Version Split
Now we describe the version split process that splits the remaining array (after optionally removing a promotable array) into a collection of arrays, each of which obeys both the minimum density constraint (L-dense) and a minimum fraction of lead elements.

We use the notion of versions that are dense in their subtrees, i.e., $v$ for which $\delta_T(v) := \delta(\lambda_T(v), v) \geq 1/3$. Intuitively, the split $\lambda_T(v)$ of a version dense in its subtree is easy to deal with from the point of view of density, but need not contain enough lead elements; on the other hand, if $v$ isn't dense, but must have a good lead ratio – in order for $u$ to not be dense, there must be many lead keys strictly descendant from $u$.

We show how to construct a split by following the 'least dense' child down the version tree, until we find a version $u$ not dense in its subtree, but all of whose children are dense in their subtrees (see Figure 1). It is not difficult to see that this always terminates; the difficult part is showing that this process finds a version $u$ with children $u_1 \ldots u_r$ and a split $\lambda(A'', \cup_i T[u_i])$ with enough lead elements and where all versions are dense. Removing the split subtree $u_1 \ldots u_i$ and recursing gives a collection of splits as required, all of which obey the required density property and all but the last of which (for which no suitable $u$ can be found) obey a lead ratio requirement. The version-split algorithm is shown in Algorithm 3.

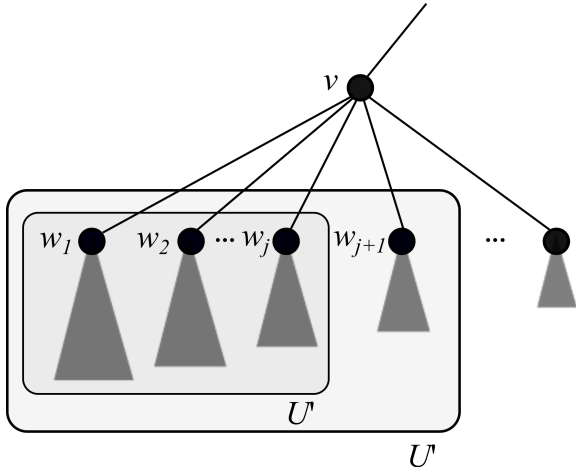

**Figure 1:** The version split process, starting from the orphan $w$. Filled nodes are dense in their subtrees.

---
**Algorithm 2** find_dense_kids$(u_1, \ldots, u_n)$
---
1: **let** $u = \arg\min \delta_T(u_i)$
2: **if** $\delta_T(u) > 1/3$ **then**
3:    **return** $[u_1, \ldots, u_n]$ sorted by *lead_below*() decreasing
4: **else**
5:    **return** find_dense_kids(children of $u$)
6: **end if**
---

The proof of the fact that version_split provides a version split with the desired properties is deferred until the full paper. Here we simply state the result:

LEMMA 2 (VERSION SPLIT). *Suppose that $(A, V)$ is a stratum obeying (P-plive)$_{l+1,V}$ and (L-no-promote). Then there is a version split of $(A, V)$, say $(A_i, V_i)$ for $i = 1 \ldots n$, such that each array satisfies (L-dense) and (L-size) for level $l$, and there is at most one index $i$ for which $lead(A_i) < |A_i|/2$.*

If $(A, V)$ also satisfies (L-live) then every split of it does (since all live elements are included), and likewise for (L-edge). It follows that version splitting $(A'', V'')$ – which necessarily has no promotable versions – results in a set of arrays all of which satisfy all of the L-* conditions necessary to stay at level $l$.

---
**Algorithm 3** version_split$(A, V, l)$
---
**Require:** $(A, V)$ is a stratum.
1: **let** $[u_1, \ldots, u_r]$ = find_dense_kids(orphans of $V$)
2: **let** split$(j) = \cup_{i \leq j} T[u_i]$
3: **for** $i = 1$ to $r - 1$ **do**
4:   **if** $|\lambda(A, \mathtt{split}(i))| > \min(2^{l+1}, 3 \cdot live(u_i))$ **then**
5:     **let** $U = \mathtt{split}(i-1)$
6:     **return** version_split$(V \setminus U, l) :: U$
7:   **end if**
8: **end for**
9: **return** [split(r)]
---

The main result of this process is the following.

LEMMA 3 (PROMOTION). *The fraction of lead elements over all output arrays after a version split is $\geq 1/39$.*

PROOF. First, we claim that under the same conditions as the version split lemma, if in addition $|A| < 2M$ and $live(v) >= M/3$ for all $v$, then the number of output strata is at most 13. Consider the arrays which obey the lead fraction constraint. Each has size at least $M/3$, since at least one version is live in it, and least half of the array is lead, so at least $M/6$ lead keys. The total number of lead keys in the array A is $\leq 2M$, since the array itself is no larger than this; it follows that there can be no more than 12 arrays obeying the lead ratio constraint, and hence no more than 13 in total.

Now, a merge at level $l$ involves at least one promoted array, which by (P-lead) contains $> 2^{l+1}/3$ lead elements. By the above, the output of the merge is at most 13 arrays of size at most $2^{l+1}$, so there are at most 39 output elements per lead element. □

### 3.5.3 Extraction

Extraction is the process of executing a version split found as in the previous section: it takes a list of disjoint version sets $\{V_i\}_i$, an iterator $it$ of $(k, v, x)$ tuples (in $k, v$ order), and outputs a set of arrays $\{\lambda(it, V_i)\}_i$ together with forward pointer arrays demoted to lower levels.

For each version set $V_i$ we create a set of output arrays $A_i^j$, one for each level. $A_i^0$ is the primary output array and will receive keys in version set $V_i$ and will end the process containing $\lambda(A'', V_i)$; the arrays $A_i^j$ for $j > 0$ will contain forward pointer samples of this array: if we are sampling with frequency $r$, then $A_i^j$ will contain a pointer to every $r^{th}$ element of $A_i^{j-1}$.

## 3.6 Lookup

A point query for $k, v$ calls query_rec$(A, k, v)$ (see Algorithm 4), where $A$ is the unique array registered to version $v$ at level 0. In general, at level $l$ we search within a lower and upper bound to find the least upper bound $(k', v', x)$ for $(k, v)$ and the associated forward pointers strictly below and weakly above this location. If $k' = k$ and $v' \preceq v$, then the least upper bound is the desired key and we return $x$ (by the ordering on versions, $v'$ must be the closest ancestor of $v$ for which a value of $k$ has been written); if $v' \not\preceq v$, then scan forwards until an ancestor is found, in which case we return it, or we reach an entry for which $k' \neq k$, in which case we recurse to the array to which forward pointers in $A$ point. The search terminates either when a suitable entry $(k, v', x)$ is found, with $v' \preceq v$, or when there are no forward pointers in $A$.

A range query query(start,end,version) is handled by performing a lookup query(start,version) (with the modification that we do not break out of arrays early as in the lookup described above). We then merge the outputs of the iterators from each of these arrays in $(k, v)$-order, with the exception that for any key $k$, we output only the first version

**Algorithm 4** query_rec($A, k, v, [lb], [ub]$)

---
**Require:** An array $A$, and optionally two locations within $A$: lower bound $lb$ and upper bound $ub$
1: **let** $loc, lb, ub = A.\text{search}(k, v, lb, ub)$
2: **let** $it = A.\text{iterate}(loc)$
3: **let** $k' = k$
4: **while** $it.\text{has\_next}()$ **and** $k' = k$ **do**
5:    **let** $(k', v', x) = it.\text{next}()$
6:    **if** $k' = k$ **and** $v' \preceq v$ **then**
7:      **return** $(k, v', x)$
8:    **end if**
9: **end while**
10: **let** $N$ be the next array of $A$
11: **if** $N \neq null$ **then**
12:    **return** query_rec($N, k, v, lb, ub$)
13: **else**
14:    **return** $null$
15: **end if**

---

ancestral to the desired version, and skip over the remaining versions.

To get the desired lookup performance, we need to modify the forward pointer construction. In the description here, FPs may not be evenly spaced within an array; in particular, for increasing inserts, all the FPs live at the start of each array and a lookup always involves a scan to the end of some arrays. This can be solved by storing, for some constant $8 < k < B$, in every $k$th element of every array a *redundant FP*, which is a copy of its two closest real FPs to the left and right. This guarantees that every element has a forward pointer within $O(1)$ blocks on either side. Space for these redundant FPs can be left in the initial output during the execution phase, and their values retrospectively updated by rescanning each output array $A_i^0$ ($A_i^j$ for $j > 0$ consists only of forward pointers, and so has no such problem).

### 3.7 Clone

On snapshot or clone of version $v$ to new descendant version $v'$, $v'$ is registered for each array $A$ which is currently registered to the parent of $v$. This does not require any I/Os.

### 3.8 Update

THEOREM 1. *The stratified doubling array performs updates to a leaf version $v$ in a cache-oblivious $O(\log N_v / B)$ amortized I/Os.*

PROOF. Assume we have at our disposal a memory buffer of size at least $B$ (recall that $B$ is not known to the algorithm). Then each array that is involved in a disk merge has size at least $B$, so a merge of some number of arrays of total size $k$ elements costs $O(k/B)$ I/Os. In the COLA [5], each element exists in exactly one array and may participate in $O(\log N)$ merges, which immediately gives the desired amortized bound. In the scheme described here, elements may exist in many arrays, and elements may participate in many merges at the same level (eg when an array at level $l$ is version split and some subarrays remain at level $l$ after the version split). Nevertheless, we shall prove the theorem using a more involved accounting argument.

We will assume that each I/O costs $\$1$ and can read or write $B$ elements. Each element $(k, v)$ inserted at version $v$ has an initial credit $\$c/B$, for some constant $c$ to be determined later. For an array $(A, W)$, recall an element $(k, v)$ a *lead element* if $v \in W$. When an array $(A, W)$ is promoted from level $l$ to $l + 1$, all its lead elements are given extra credit $\$c/B$. Assume for now that this is sufficient to pay for all I/O operations. By the level condition (L-live), all arrays $(A, W)$ with $v \in W$ must live in levels $l \leq \lg N_v + O(1)$. This implies that the total charge to element (k,v) is $O(\log N_v/B)$, since $(k, v)$ appears as a lead element in exactly one array, is only charged when it appears as a lead element (hence it can only be charged at those levels where $v \in W$), and lead elements can never be demoted (so it is charged at most once per promotion). It now remains to prove the assumption.

LEMMA 4. *For $c > 45$, the credit of every element $(k, v)$ is $\geq 0$.*

PROOF. Each array is either dead or alive. An array at level $l$ is alive if it enters level $l$ by being promoted from level $l - 1$, and becomes dead at level $l$ if it enters level $l$ as a result of a version split. Consider a merge at level $l$. The algorithm guarantees that at least one of these arrays is alive. We will charge the entire cost of the merge to the lead elements in the alive arrays participating in it (if there is more than one such array, divide the cost equally between their lead elements).

Consider a merge at level $l$. It involves $O(1)$ passes over $O(1)$ input arrays, at least one of which is alive (otherwise the promotion would not have been triggered), followed by a version split which produces some output arrays. The Promotion Lemma implies that for $c > 39$, the lead elements in the alive array can pay for all the I/Os involved in producing the output arrays. Since the input array has just been promoted, by (P-lead), it has at least $2^{l+1}/3$ lead elements. The total input size is at most $2^{l+2}$, so to perform the input and output passes, it suffices to ensure that $c > 39 + 6 = 45$. □

The theorem follows since the lead elements of alive arrays can pay for all the I/Os involved in merging and splitting. □

### 3.9 Lookup

For large range queries (that retrieve a constant fraction $Z = \Omega(N_v)$ of the live keys of some version $v$), the density property of the arrays immediately gives an asymptotically optimal bound of $O(\log N_v + Z/B)$. For much smaller range queries, the worst-case performance may be the same as for a point query. We now prove the amortized bound, which applies to smaller queries.

THEOREM 2. *A range query at version $v$ costs $O(\log N_v + Z/B)$ amortized I/Os.*

PROOF. We first consider just point queries, and amortize the cost of $lookup(k, v)$ over all keys live at $v$. Let $l(k, v)$ be the cost of $lookup(k, v)$, then the amortized cost is given by $\sum_k l(k, v)/N_v$.

For an array $A_i$, let $l(k,v,A_i)$ be the number of I/Os used in examining elements in $A_i$ for lookup(k,v). The idea is that since the elements of $A_i$ are $(k,v)$-ordered, the parts of $A_i$ examined by each key lookup for version $v$ are disjoint, hence $\sum_k l(k,v,A_i) \leq |A_i|$. We have the following:

$$\begin{aligned}\frac{\sum_k l(k,v)}{N_v} &= \frac{\sum_i \sum_k (l(k,v,A_i) + O(1))}{N_v} \\ &\leq \frac{\sum_i |A_i| + \sum_k \sum_i O(1)}{N_v} \\ &= \frac{O(N_v) + O(N_v \log N_v)}{N_v} = O(\log N_v).\end{aligned}$$

The $O(1)$ additive term is due to finding and following FPs between arrays, which follows from the redundant FP construction. The second inequality follows since there are $O(\log N_v)$ arrays examined during the searches for all keys in some fixed version $v$, and the first term follows from the density property and the geometrically increasing array sizes.

A range query incurs the same initial lookup cost (to locate the starting points of the iterators in each array), and then the subsequent scan can be analysed in the same way as above. We can also deamortize the point query bound to worst-case $O(\log N_v)$ by embedding a small lookup structure into the part of each array defined by each key $k$. The proofs will appear in the full version of the paper. □

## 3.10 Space
We can prove that the structure has asymptotically optimal space requirements.

THEOREM 3. *The structure uses $O(N)$ external memory space to store $N$ elements, regardless of the version tree.*

PROOF. It is easy to see that there are exactly $N$ lead elements. Whenever an array containing $k$ lead elements is promoted, Theorem 1 established that at most $O(k)$ space is used and possibly never freed (used by dead arrays). Each lead element gets promoted exactly once and by the promotion conditions, the number of lead elements must double between successive promotions. Thus the total space is at most $O(\sum_{i>0} N/2^i) = O(N)$. □

## 3.11 Deamortized updates
A single update in the SDA algorithm may trigger a cascade of merges, in the worst case requiring $\Omega(N/B)$ I/Os. With some effort, we can deamortize the merge and insert processes, such that lookups are valid at every point in time, and an insert to version $v$ takes worst-case $O(\log N_v)$ IOs, and amortized $O(\frac{\log N_v}{B})$ IOs as before.

## 3.12 Cache-aware tradeoffs
Examining the level conditions in Section 3.3.1, arrays approximately double in size between successive levels. Similarly to Bender et al. [5], we can obtain a range of 'cache-aware' query/update tradeoffs by selecting a 'growth factor' $g = B^\varepsilon \geq 2$. Some work is needed to ensure the version splitting process, and we can obtain the following results, the proof of which is deferred to the full paper.

THEOREM 4. *The data structure can support updates to version $v$ in amortized $O(\frac{\log_B N_v}{\varepsilon B^{1-\varepsilon}})$ I/Os, and range queries of size $Z$ in amortized $O(\frac{\log_B N_v}{\varepsilon} + \frac{Z}{B})$ I/Os.*

## 4. EXPERIMENTAL RESULTS
We implemented a prototype of the versioned B-tree and the cache-oblivious versioned B-tree in OCaml, an efficient compiled functional language. The machine had 1GB RAM (but we restricted 256MB to be available for the buffer cache in the tests), a 2GHz AMD Athlon 64 processor (although our implementation was only single-threaded), a 500GB SATA disk and an Intel X25-M SSD. We used a block size of 32KB for the B-tree on disk and 4KB on SSD. The disk can perform $\approx 50$ such I/Os/s; by contrast, the SSD can perform 35,000 random 4KB reads/s but must write in blocks of 512KB; however some buffering tricks in the firmware allow writes of 4KB blocks. [2] There was no such tuning to be done for the cache-oblivious structure.

We started with a single root version and inserted random 100 byte key-value pairs to random leaf versions, and periodically performed range queries of size 10,000 at a random version. Every 100,000 insertions, we create a new version as follows: with probability 1/3 we clone a random leaf version and w.p. 2/3 we clone a random internal node of the version tree. This aims to keep to the version tree 'balanced' in the sense that there are roughly twice as many internal nodes as leaves.

The figures show results for the versioned B-tree (btree), the SDA (strat-DA) and, for comparison, the SDA where we forbid any version splitting; hence there is a single array at each level. Figure 2 shows the insertion performance on the disk. As expected, the B-tree performance degrades rapidly when the dataset exceeds internal memory available. Figures 3 and 4 show range query performance on disk and the SSD. The SDA beats the B-tree and DA by a factor of more than 10 on both disk and SSD, while the versioned B-tree beats the non-version-split DA on SSD, likely due to excellent random read performance (on disk, the overhead of scanning over irrelevant versions in the DA appears to be low compared to the overhead of performing random reads).

**Acknowledgments.** We would like to thank Richard Low, Grzegorz Milos, Tim Moreton, Andrew Suffield and John Wilkes for helpful discussions, comments and suggestions.

---
[2]The datasheet claims 3,300 random 4K writes/s and large sequential writes at over 150MB/s, so there is a large advantage in being able to write sequentially.

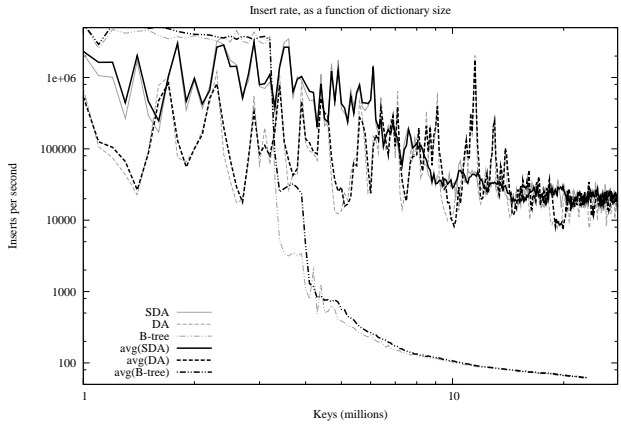

**Figure 2: insert performance with 1000 versions on disk.**

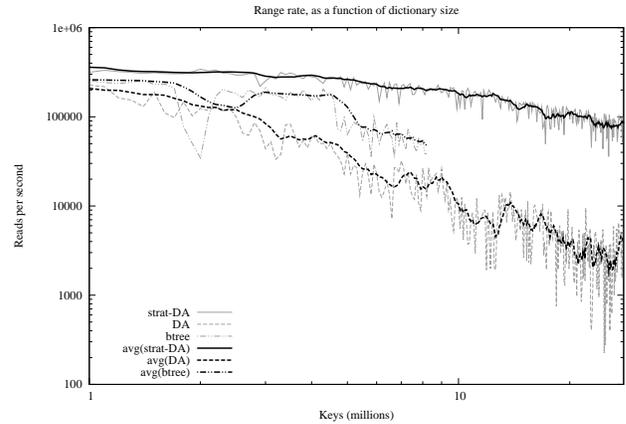

**Figure 4: range query performance with 1000 versions on ssd.**

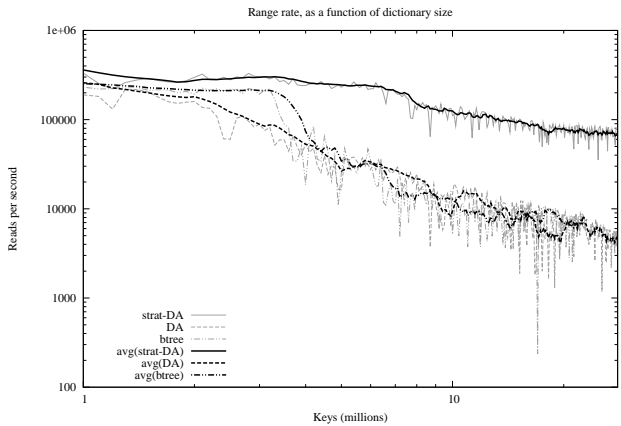

**Figure 3: range query performance with 1000 versions on disk.**

# APPENDIX
## A. PROOFS
### A.1 Version Split Lemma

We prove the Version Split Lemma, which is crucial for maintaining density within levels, and for bounding the amount of merge work we need to do. We first demonstrate that there is some redundancy in the P-* conditions:

LEMMA 5. *If $(A,V)$ is a stratum for which $(P\text{-plive})_{l,V}$ and $(L\text{-no-promote})$ hold, then $(P\text{-live})_{l+1,v} \implies (P\text{-max-size})_{l,\lambda(A,TVv)}$, i.e.,*

$$live(v) \geq 2^{l+1}/3 \implies |\lambda(A,TVv)| < 2^{l+1}. \qquad (3)$$

*In particular, all such versions are dense in their subtrees.*

PROOF. Suppose by way of contradiction that there exists a version $w$ such that $live(w) \geq 2^{l+1}/3$ and $|\lambda(A,TVw)| \geq M$. Let $v$ be the oldest ancestor of $w$ in $V$ for which $live(v) \geq 2^{l+1}/3$. Since $v$ is an ancestor of $w$, $|\lambda(A,TVv)| \geq |\lambda(A,TVw)|$ and so (P-live) and (P-min-size) both hold for $v$. By (L-no-promote), $lead\_below(v) < 2^{l+2}/3$. Whether $v$ is an orphan of $V$ (in which case $(P\text{-plive})_{l,V}$ is needed) or not, it must be the case that $live(parent(v)) < 2^{l+1}/3$, so $|\lambda(A,TVv)| \leq live(parent(v)) + lead\_below(v) < 2^{l+1}$, in contradiction to (P-min-size). □

The following lemma forms the basis of the Version Split Lemma.

LEMMA 6. *Suppose $(A,W)$ is a stratum in level $l$ such that, for $M = 2^{l+1}$ we have*

1. *(cD) $live(v) < M/3$ for $v = parent(W)$ and all versions $v \in W$ that are not dense in their subtrees; and*
2. *(cS) $|\lambda(A,TVv)| < M$ for all $v \in W$ such that $v$ is dense in its subtree.*

*Then `version_split`$(W,l)$ gives a version split $W_i$ of $(A,W)$ such that the associated extracted arrays $A_i = \lambda(A,W_i)$ satisfy:*

1. *(vS) $|A_i| \leq M$ for all $i$;*
2. *(vD) $A_i$ is dense for all versions in $W_i$;*
3. *(vL) $\frac{lead(A_i)}{|A_i|} \geq \frac{1}{2}$ for all but at most one $i$.*

PROOF. Proof is by induction on the size of $W$. Consider a pass through the `version_split` algorithm.

The subroutine `find_dense_kids` returns the children $u_1 \ldots u_r$ of some version $u$, such that all $u_i$ are dense in their subtrees; the children are ordered decreasing by $lead\_below$. It returns the orphans of $W$ iff all orphans of $W$ are dense in their own subtree; in this case $u = parent(W)$ and it is not known whether $u$ is dense in $A$ or not; in all other cases $v$ is not dense in its subtree.

As a particular case of the density of the $u_i$, $u_1$ is dense in its subtree, which from (cS) means that $|\lambda(A,TVu_1)| < M$; density implies that $|\lambda(A,TVu_1)| < 3live(u_1)$. Therefore the size test on line 4 of the algorithm must evaluate to false, and we never split at $u_1$.

If the test evaluates to true for $i > 1$ then, since it was false at $i-1$, the array $U$ constructed in line 7 must satisfy $|\lambda(A,U)| \leq M$ (vS), and is dense for all versions (vD). This also holds if the loop exits without having found an upper bound $i$: $U = \text{split}(r)$ is still small enough and dense for all versions. In this case we are done since the version split list has a single entry, so (vL) holds. This is the base case for induction, since it exhausts $W$.

In the former case, where $\text{split}(i)$ fails the size test, the lemma will proved by induction so long as we can establish firstly that the conditions of the lemma still hold for $W \setminus U$, and secondly that (vL) holds for $U$, for which it suffices to prove that $live(u) < lead(U)$, since $|\lambda(A,U)| \leq live(u) + lead(U)$.

The condition (cD) is trivially maintained, since the set of versions not dense in their subtrees can only shrink. For (cS), we have to check that if $v$ is a version in $W \setminus U$, not dense in $\lambda(A,TWv)$, but made dense in $\lambda(A,TW \setminus Uv)$ by the removal of keys with versions strictly descendant from $v$, then $|\lambda(A,TW \setminus Uv)| < M$ (to maintain (cS)). However, since such a version is not dense prior to the removal of $A_i$, it follows from (cD) that $live(v) < M/3$, and so post-hoc density implies $|\lambda(A,TW \setminus Uv)| < 3 \cdot live(v) < M$ as required for (cS).

We now return to the case where there is an $i > 1$ such that $U = \text{split}(i-1)$ passes the size test, but $U' = \text{split}(i)$ fails. Failure at $i$ implies either that there is a $j \leq i$ such that $u_j$ is not dense in $U'$, or that $|U'| > M$. In either case, since $u_k$ are ordered by $lead\_below()$ decreasing,

$$\begin{aligned}
lead(U') &= \sum_{k=1}^{i} lead\_below(u_k) \\
&\leq \frac{i}{i-1} \sum_{k=1}^{i-1} lead\_below(u_k) \\
&\leq 2lead(U) \\
\Rightarrow lead(U) &\geq \frac{1}{2} lead(U') \qquad (4)
\end{aligned}$$

If $|U'| > M$ then we can use (cD) directly: $live(v) < M/3$, and so $lead(U') \geq |U'| - live(v) > 2M/3$, which implies $lead(U) > M/3$ from (4), and so (vL) holds for $U$. If on the other hand $u_j$ is not dense in $U'$ then

$$\begin{aligned}
3live(v) &\leq 3live(u_j) \\
&< |U'| \\
&\leq live(v) + lead(U') \\
\Rightarrow 2live(v) &< lead(U') \\
&\leq 2lead(U).
\end{aligned}$$

In either case $live(v) < lead(U)$ and therefore (vL) holds. □

Now we apply this lemma to prove the version split lemma.

PROOF OF VERSION SPLIT LEMMA. To prove the lemma we must establish that (cD) and (cS) are a consequence of (P-plive)$_{l+1,V}$ and (L-no-promote). (P-plive)$_{l+1,V}$ is exactly the first clause of (cD), namely $live(parent(V)) < M/3$. Consider any version $v$. From Lemma 5, $v$ is dense in its subtree whenever $live(v) \geq M/3$, the contrapositive of which is that whenever $v$ is not dense in its subtree, $live(v) < M/3$, i.e. (cD). On the other hand, if $v$ is dense in its subtree and $|\lambda(A, TVv)| \geq M$ then by the definition of density, $live(v) \geq M/3$, a contradiction to Lemma 5. Thus, we must have $|\lambda(A, TVv)| < M$, proving (cS). □